\documentclass[twocolumn,secnumarabic,amssymb, aps,prb,reprint,longbibliography]{revtex4-1}
\usepackage{graphicx}
\usepackage{color}
\usepackage{dcolumn}
\usepackage{bm}
\usepackage{ucs}
\usepackage{physics}
\usepackage{lineno}
\usepackage{ulem}
\usepackage[utf8]{inputenc}
\usepackage[T1]{fontenc}
\usepackage{units}
\usepackage{multirow}
\usepackage{booktabs}
\usepackage{placeins}
\usepackage{mathtools}
\usepackage{hyperref}
\usepackage{float}

\begin{document}

\title{Entanglement between quasiparticles in superconducting islands mediated by a single spin}

\author{Juan Carlos Estrada Salda\~{n}a$^{1}$}
\email{juan.saldana@nbi.ku.dk}
\author{Alexandros Vekris$^{1,2}$}
\author{Luka Pave\v{s}i\v{c}$^{3,4}$}
\author{{Rok \v{Z}itko}$^{3,4}$}
\author{Kasper Grove-Rasmussen$^{1}$}
\author{Jesper Nyg{\aa}rd$^{1}$}
\email{nygard@nbi.ku.dk}

\affiliation{$^{1}$Center for Quantum Devices, Niels Bohr Institute, University of Copenhagen, 2100 Copenhagen, Denmark}
\affiliation{$^{2}$Sino-Danish College (SDC), University of Chinese Academy of Sciences}
\affiliation{$^{3}$Jo\v{z}ef Stefan Institute, Jamova 39, SI-1000 Ljubljana, Slovenia}
\affiliation{$^{4}$Faculty of Mathematics and Physics, University of Ljubljana, Jadranska 19, SI-1000 Ljubljana, Slovenia}

\flushbottom
\maketitle

\newcommand{\red}[1]{{\color{red}#1}}

\thispagestyle{empty}

\textbf{Condensed matter is composed of a small set of identical units, yet it shows an immense range of behaviour. Recently, an array of cold atoms was used to generate long-range quantum entanglement, a property of topological matter~\cite{Semeghini2021Dec}. Another approach to strong non-local correlations~\cite{Ranni2021Nov,Wang2022, Bordoloi2022} employs the macroscopic coherence of superconductors~\cite{Hofstetter2009Oct,HerrmannPRL2010,Matsuo2022}. Impurity spins in superconductors are thought to be unamenable to the formation of long-range spin entanglement because each spin tends to be screened by binding to a quasiparticle from the superconductor to form a local singlet~\cite{DeaconPRL2009,lee2014spin,Saldana2022Jan,BagerbosPRXQ2022}. Here we demonstrate that it is possible to attach a second quasiparticle to the spin, overscreening it into a doublet state carrying ferromagnetic correlations between two quasiparticles over a micrometer distance. To demonstrate this effect, which is strongest for equal binding, we symmetrically couple the spin of a quantum dot to two ultrasmall superconducting islands.
The overscreened state requires sufficiently large Coulomb repulsion and exchange binding to become well defined.
We predict that this state will carry long-range correlations for an alternating chain of quantum dots and superconducting islands, opening a new route to controllable large-scale entanglement in the solid state.}

\begin{figure}[t!]
\centering
\includegraphics[width=1\linewidth]{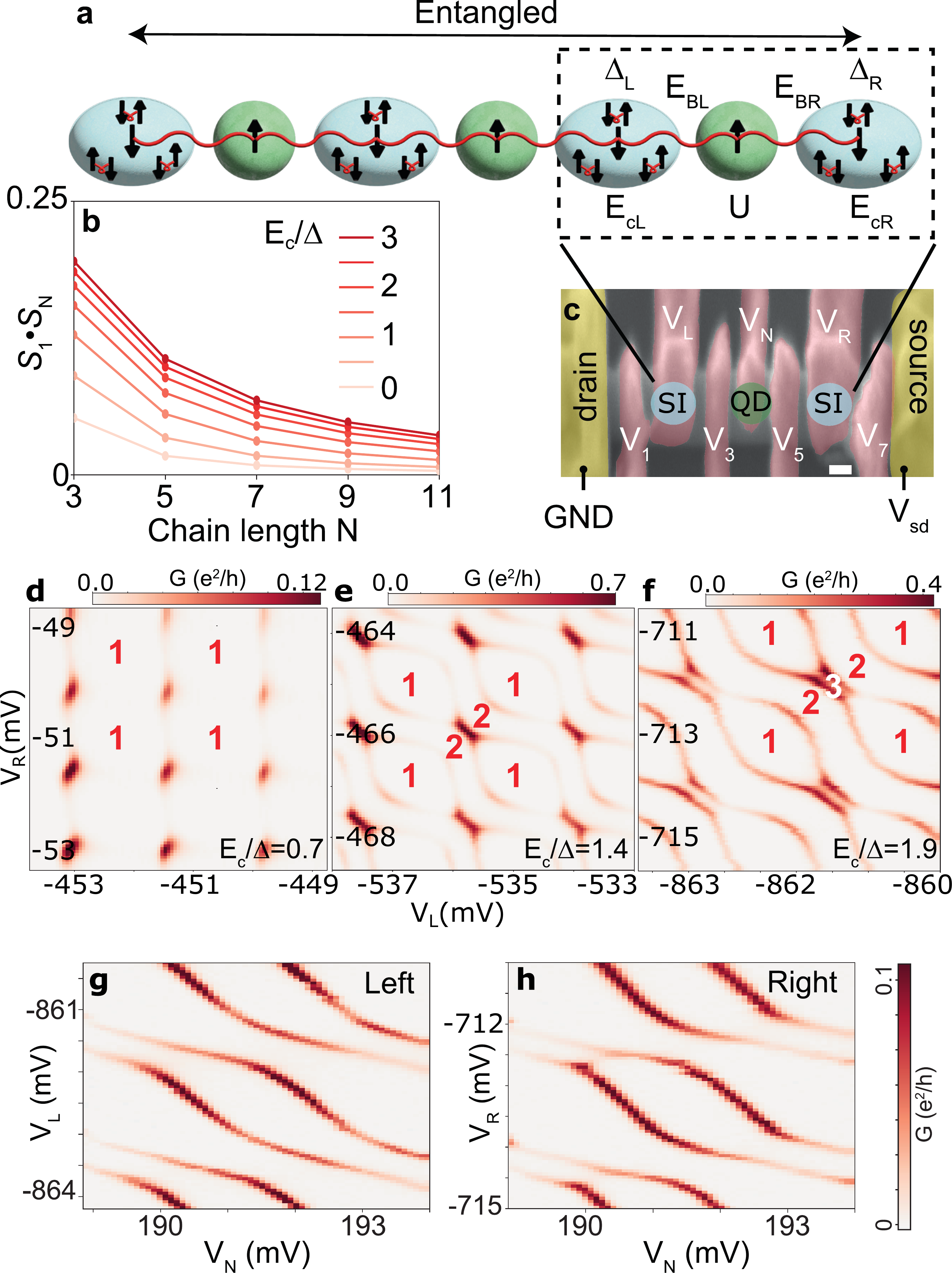}
\caption{\textbf{Chain of Bogoliubov quasiparticles entangled by electron spins.} \textbf{a} Alternating chain of superconducting islands (SI) and quantum dots (QD). Each SI carries one Coulomb-blockaded quasiparticle, each QD one Coulomb-blockaded spin. In this arrangement, the Yu-Shiba-Rusinov interaction antiferromagnetically binds neighbouring local moments (LMs) within the chain. For equal bindings each QD spin is overscreened by two adjacent quasiparticles. This state has long-range ferromagnetic correlations between every second element of the extended doublet ground state. \textbf{b} Calculated correlations between the spins at the ends of the chain versus chain length for a range of Coulomb repulsions $E_\mathrm{c}$.
\textbf{c} Device used to realize a three-element chain consisting of a QD coupled to two SIs. Scale bar is 100 nm. 
\textbf{d,e,f} Zero-bias conductance $G$ versus $V_\mathrm{L}$, $V_\mathrm{R}$ with increasing $E_c/\Delta$, illustrating the development of the overscreened doublet regime with 3 LMs. Numbers in the diagram indicate the total number of LMs in the device.
\textbf{g,h} Zero-bias conductance $G$ versus $V_\mathrm{N}$, and (\textbf{g}) $V_\mathrm{L}$ or (\textbf{h}) $V_\mathrm{R}$. The resemblance in the conductance patterns in the two diagrams reflects the high degree of left-right symmetry of the device parameters in this gate configuration, with the differences in $E_\mathrm{c}/\Delta$ and binding energy $E_\mathrm{B}$ being 8\% and 14\%.
}
\label{Fig1}    
\end{figure}

\begin{figure}[t!]
\centering
\includegraphics[width=1\linewidth]{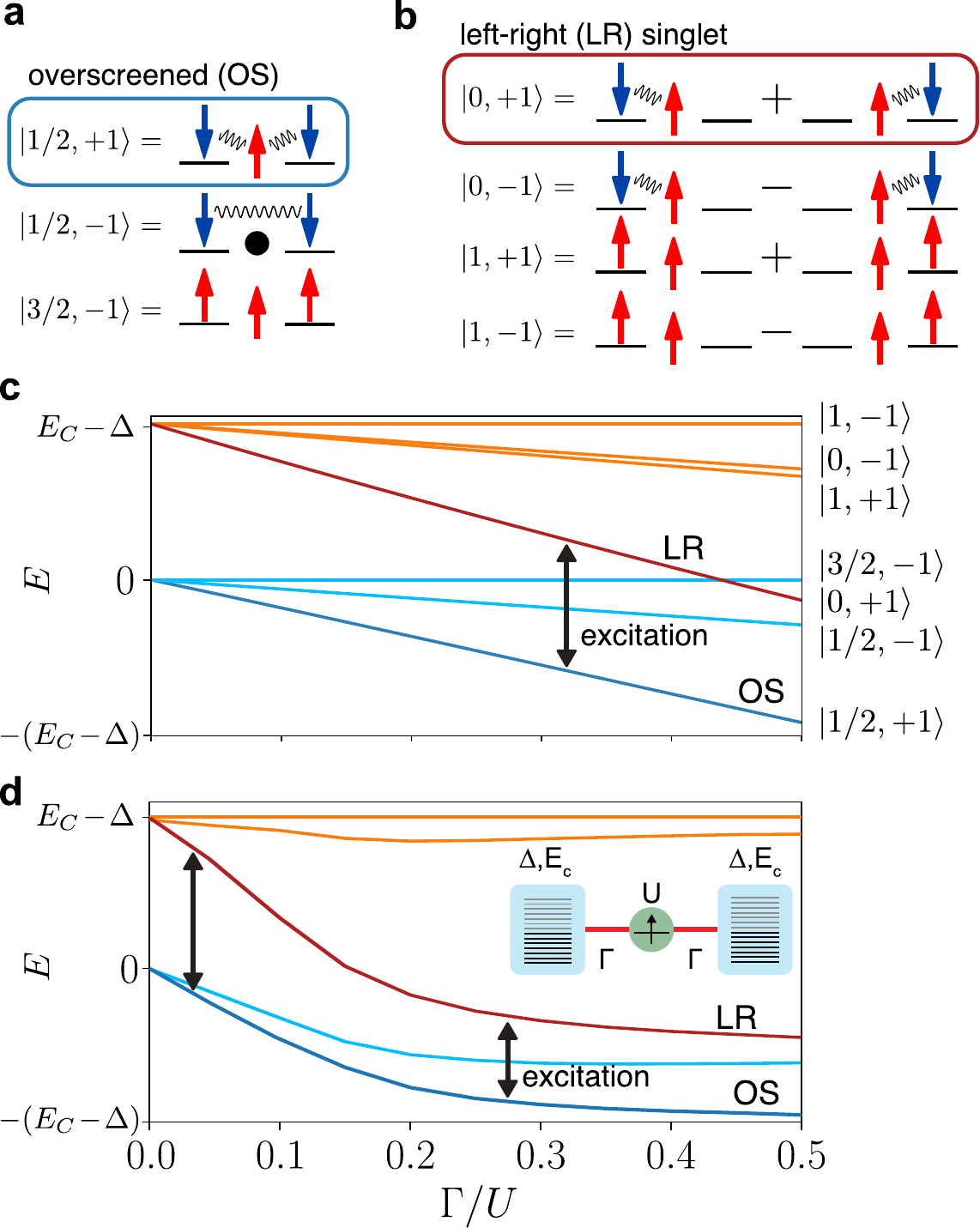}
\caption{\textbf{Manifold of states in the overscreened regime.} 
\textbf{a, b} Low-energy states 
with odd (\textbf{a}) and even (\textbf{b}) occupation, labelled by total spin $S$ and space-inversion parity $P$, where $P=\pm 1$ denotes a gerade/ungerade superposition. A squiggly line between two spins symbolizes an entangled (singlet) configuration, black circle is a decoupled local moment. 
\textbf{c} Energy spectrum vs. hybridisation $\Gamma$ in the single-level approximation for $U=100 \Delta$, $E_c = 5 \Delta$.
States are labelled as in (\textbf{a}) and (\textbf{b}). The arrow indicates the transition energy between the lowest doublet and singlet states, plotted in Fig.\ref{Fig3}e,f and discussed in text.
\textbf{d} Energy spectrum of the full model for $U=10\Delta$, $E_c=2\Delta$.
Blue states correspond to an odd, red and orange to an even number of particles. 
The QD-SI coupling is quantified by the hybridisation strength $\Gamma$, which determines the binding energies $E_B$ (see Extended Data Fig.~\ref{Ext10}).
\textbf{Inset} Sketch of the full model; see Methods for details.
}
\label{Fig2}    
\end{figure}

\begin{figure*}[t!]
\centering
\includegraphics[width=1\linewidth]{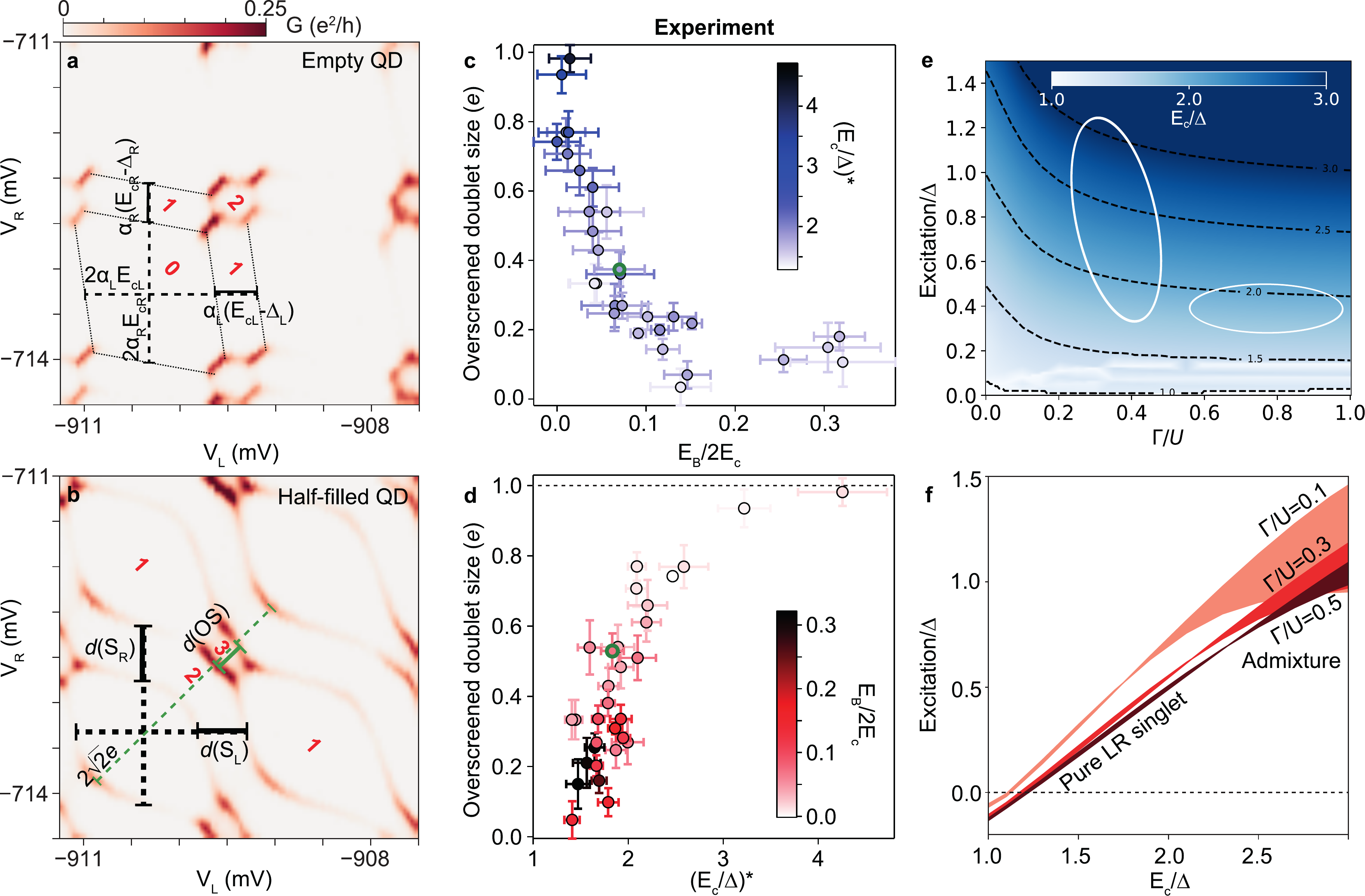}
\caption{\textbf{Binding of two quasiparticles to one spin.} \textbf{a,b} Stability diagrams presented as zero-bias $G$ versus $V_\mathrm{L}$ and $V_\mathrm{R}$ for \textbf{a} empty QD, \textbf{b} QD filled with one LM. The total number of LMs in the device is indicated in red numbers. Solid bars in \textbf{(a)} normalized by the dashed bars provide estimates for the $E_\mathrm{c}/\Delta$ ratios independent of the gate-to-energy conversion factors, denoted as $(E_\mathrm{c}/\Delta)^*$.
Solid horizontal and vertical bars in \textbf{(b)} minus those in \textbf{(a)}, normalized by the dashed bars, provide $E_\mathrm{B}/2E_\mathrm{c}$. The diagonal solid bar of length $d(\mathrm{OS})$ in \textbf{(b)} measures the size of the OS doublet when it is the ground state. The diagonal dashed bar measures the gate voltage needed to add two electrons in each SI, and we use it as a normalization factor for the size of the OS region.
\textbf{c} Normalized OS region size in 26 QD shells, $2\sqrt{2}d(\mathrm{OS})$, against $E_\mathrm{B}/2E_\mathrm{c}$. $(E_\mathrm{c}/\Delta)^* = 1.3 - 4.2$, given by the color scale. Left-right differences in $E_\mathrm{B}$ and $E_\mathrm{c}/\Delta$ are 0-30\%. $U/\Delta=2-4$ measured from Coulomb diamonds spectroscopy.
\textbf{d} Data from (\textbf{c}) plotted against $(E_\mathrm{c}/\Delta)^*$. $E_\mathrm{B}/2E_\mathrm{c} = 0 - 0.32$, given by the color scale. The green circle indicates the data point extracted from (\textbf{a},\textbf{b}).
Error bars are based on the full width at half-maximum of the conductance lines used in the extraction of the data. 
\textbf{e} Calculated doublet$\to$singlet excitation energy versus $\Gamma/U$ and $E_c$ in the center of the 3 LM sector. The white ellipsis indicate the approximate position of the two clusters shown in \textbf{c}.
\textbf{f} Calculated doublet$\to$singlet excitation energy versus $E_\mathrm{c}/\Delta$ for different $\Gamma/U$. Bands in (\textbf{e},\textbf{f}) indicate the parameter range $U/\Delta=2-4$ used in the calculations.}
\label{Fig3}
\end{figure*}
The superconductor-semiconductor hybrid platform 
is excellent for exploring non-local properties of sub-gap states~\cite{MenardPRL2020}, hybrid-based quantum bits~\cite{Tosi2019Jan,HaysScience2021,Pita-VidalarXiv2022},  
and non-local processes \cite{Ranni2021Nov,Wang2022,Bordoloi2022,Kurtossy2021} for topological \cite{Dvir2022, Wu2021} and non-topological chains \cite{deJong2022}. An alternating superconducting island (SI)-quantum dot (QD) chain is schematized in Fig.~\ref{Fig1}a. All SIs have equal charging energy $E_\mathrm{c}$ and superconducting energy gap $\Delta$ \cite{
Averin1992,vonDelft2001,Higginbotham2015Sep,Albrecht2016Mar,Shen2018Nov}, and all QDs have equal Coulomb repulsion $U$ \cite{
de2010hybrid,Prada2020Oct}. A Bogoliubov quasiparticle occupies each SI, which is facilitated for $E_\mathrm{c}>\Delta$ \cite{Averin1992,vonDelft2001,theory,Higginbotham2015Sep,Saldana2022Jan}. The exchange interaction binds each quasiparticle to an antiparallel spin in the adjacent QD~\cite{
choi2004josephson,oguri2004josephson,meden2019anderson,DeaconPRL2009,lee2014spin,jellinggaard2016tuning,Grove-Rasmussen2018,Saldana2022Jan,BagerbosPRXQ2022}. When the binding energy $E_\mathrm{B}$ is equal among all SI-QD pairs, the ground state (GS) of the whole system becomes a doublet with long-range ferromagnetic correlations between the end spins, as shown in Fig.~\ref{Fig1}b. The correlations decrease with the chain length, but are recovered by increasing $E_\mathrm{c}/\Delta$ which stabilizes the local moments (LM) in the SIs. 
The patterns of spin correlations in chains with an odd number of elements are similar for chains of different lengths if $E_\mathrm{c}/\Delta$ is sizable (see Extended Data Fig.~\ref{Ext9}). 
Therefore, a three-element chain (box in Fig.~\ref{Fig1}a) is representative of the longer chain, and it is the unit investigated here (Fig.~\ref{Fig1}c).

\begin{figure*}
\centering
\includegraphics[width=1\linewidth]{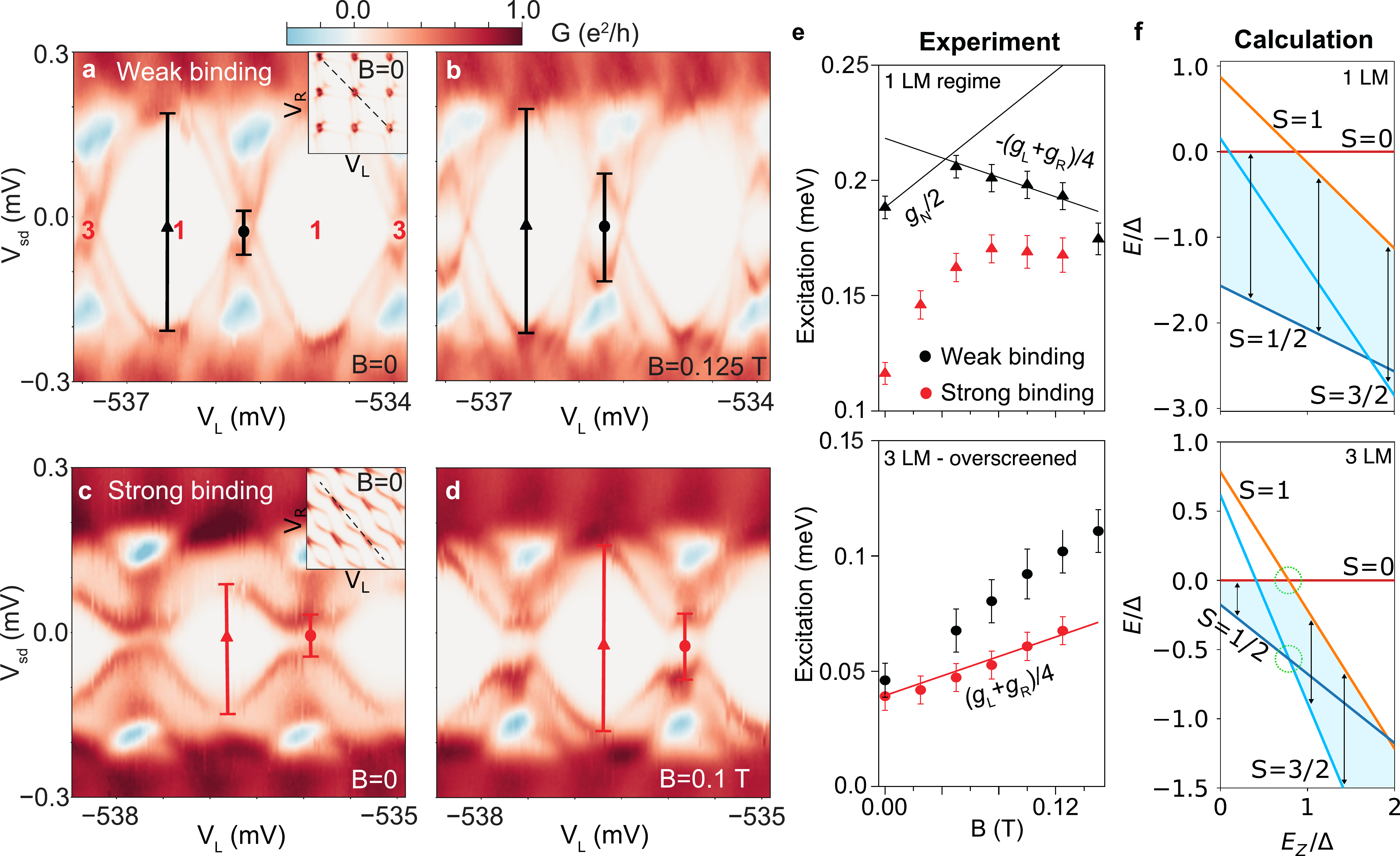}
\caption{\textbf{Polarizing two bound Bogoliubov quasiparticles with magnetic field.} \textbf{a-d} Bias spectra for approximately left-right symmetric parameters, with (\textbf{a,b}) $E_\mathrm{B}/2E_\mathrm{c}=0.04$ (weak binding), $E_\mathrm{c}/\Delta=1.45$ and (\textbf{c,d}) $E_\mathrm{B}/2E_\mathrm{c}=0.32$ (strong binding), $E_\mathrm{c}/\Delta=1.65$, recorded at different $B$ indicated on each plot. The gate sweep, indicated by a dashed line in the inset stability diagrams in (\textbf{a,c}) for $B=0$, alternates the total number of LMs between 1 and 3, while keeping a LM fixed in the QD. At $B=0$, the GS is a doublet and the first excited state a singlet. These states are overran by higher-spin states at larger $B$. \textbf{e} $B$ dependence of excitation energy for weak (black symbols) and strong (red symbols) binding, in the regime with either 1 LM (top plot) or 3 LM (bottom plot); in the latter case the GS is the OS state. 
The lines indicate theory estimates (see text for details).
\textbf{f} Calculation of the spectrum versus the Zeeman energy of the QD, $E_\mathrm{z}$, assuming equal $g$-factors in all parts of the device, in the regime with either 1 LM (top plot) or 3 LM (bottom plot). States are labeled with total spin $S$. Black arrows indicate the energy difference between the ground state in the sector with odd and even number of particles; this is the excitation energy shown in \textbf{e}. Green circles in the bottom plot indicate the singlet-triplet and doublet-quadruplet crossings. $U/\Delta=4$, $E_c/\Delta = 1.5$ and $\Gamma/U=0.3$. }
\label{Fig5}
\end{figure*}

An InAs semiconductor nanowire hosts a gate-defined QD whose occupation is tuned with the top-gate voltage $V_\mathrm{N}$. The binding energies $E_\mathrm{BL}$, $E_\mathrm{BR}$ to two Al SIs are controlled by $V_\mathrm{3}$ and $V_\mathrm{5}$. 
$E_\mathrm{cL}/\Delta_\mathrm{L}$ and $E_\mathrm{cR}/\Delta_\mathrm{R}$ are individually tuned by coarse changes in $V_\mathrm{L}$ and $V_\mathrm{R}$.
This makes it possible to tune the system to the regime of $E_\mathrm{cL}/\Delta_\mathrm{L}>1$ and $E_\mathrm{cR}/\Delta_\mathrm{R}>1$ where the occupations of the SIs can be accurately tuned by further fine adjustments of $V_\mathrm{L}$ and $V_\mathrm{R}$
(see Fig.~\ref{Fig1}d,e,f and Methods for details). Standard lock-in techniques are used to obtain the differential conductance $G$, from which we extract excitation energies. 
We tune the device to left-right symmetric $E_\mathrm{c}/\Delta$ and $E_\mathrm{B}$ by comparing pairs of zero-bias $G$ diagrams, of which an example is shown in Figs.~\ref{Fig1}g,h. 
The high symmetry that we are able to achieve relies on high device tunability and on having designed the SIs to be nominally identical by crystal growth and lithography, an advantage over gate-defined QD chains~\cite{Hensgens2017Aug}. 

When $V_L$ and $V_R$ are fine tuned to odd occupation, it is favourable for one Bogoliubov quasiparticle to occupy each SI. If furthermore $V_N$ is tuned so that a QD is occupied by a single electron, in total up to three unpaired spins can be confined in the device.
The eigenstates that are key for understanding the device behaviour are sketched in Fig.~\ref{Fig2}a and b (the complete set is presented in Extended Data Fig.~\ref{Ext0}).
The states in the three-particle subspace (Fig.~\ref{Fig2}a) have one quasiparticle in each SI,
while those in the two-particle subspace (Fig.~\ref{Fig2}b) only one, with the two spins combined into gerade and ungerade superpositions on three sites. In the weak-coupling limit, the energy difference between the two subspaces is $E_c-\Delta$, which accounts for the energy cost of splitting one further Cooper pair and the gain in the charging energy for single occupancy of the SI.

Fig.~\ref{Fig2}c shows the energy spectra as the coupling to the QD is increased, computed in the highly simplified single-level approximation where each SI is represented by one energy level. 
The state $\ket{0, +1}$ represents the well-known YSR singlet \cite{yoshioka2000,meden2019anderson}, here formed as a gerade superposition of contributions from each SI \cite{Pavesic2021Oct}. Due to its symmetric nature we name this state the left-right (LR) singlet. 
For $E_c=0$, the doublet GS would consist of SIs in the BCS ground state and a decoupled LM on the QD \cite{fabrizio2017non,meden2019anderson}. In contrast to this, the presence of 3 LMs in our system (strongly enforced by large $E_c$ \cite{theory,Pavesic2021Oct}) leads to a very different state. The quasiparticles from both SIs screen the spin in the QD, resulting in an overscreened (OS) state \cite{Cox1996Nov,Oreg2003Apr,affleck2005non,Potok2007Mar,fabrizio2017non} with long-range ferromagnetic spin correlations between the two quasiparticles, which is the main focus of this work.

To establish this account of the eigenstates, we introduce a more elaborate model of the chain where each SI consists of several hundred levels, as sketched in the inset of Fig.~\ref{Fig2}d. Using the density matrix renormalization group (DMRG) method we calculate the GS and a number of excited states \cite{theory, Pavesic2021Oct}. 
A key result is that the OS doublet is, in fact, closely related to the LR singlet. In particular, their binding energies are essentially the same:
the OS-LR excitation energy saturates with increasing QD-SI coupling to a value that depends only on the energy difference between the states in the zero-coupling limit, which is proportional to $E_c$. 
This behaviour is only found for $E_c>\Delta$. In the opposite case the increasing coupling induces the well-known doublet-singlet phase transition \cite{theory}.
We thus use this unique property as the experimental check for the presence of the OS state. It is important to acknowledge that the eigenstates of the system are not simply the pure states schematically represented in Fig.~\ref{Fig2}a,b, but rather complicated superpositions of them. For example, the doublet-singlet excitation energy in Fig.~\ref{Fig2}d  first slightly decreases at small coupling because the doublet GS in this regime is found to be a superposition of the decoupled state (second line in Fig.~\ref{Fig2}a) and the OS state; the crossover to a pure OS state occurs only at large coupling $\Gamma$. When the OS state becomes the dominant component the excitation energy indeed saturates. 

The competition between the various ground states is experimentally investigated by sweeping the gate voltages that control the number of LMs and their distribution within the device. We examine how the size of the OS doublet domain is affected by the parameter variation. Figs.~\ref{Fig3}a,b show examples of zero-bias $G$ stability diagrams versus $V_\mathrm{L}$, $V_\mathrm{R}$ when the QD is empty (a) and when it is filled with one LM (b) to illustrate how the OS doublet size and the parameters $(E_\mathrm{c}/\Delta)^*$ and $E_\mathrm{B}$ are gauged. Here, the asterisk indicates that the quantity is extracted from the experiment and thus slightly different from the bare $E_\mathrm{c}/\Delta$ (see caption to Fig.~\ref{Fig3} and Extended Data Fig.~\ref{Ext10}); such parameter extraction is made possible by the periodicity of the diagrams (period given by dashed lines).
The stability diagrams are namely periodic in the $V_L$, $V_R$ plane, the periodicity determined by the gate voltage required to add two electrons in the $E_\mathrm{c}/\Delta \to \infty$ limit. In order to relate the results from different shells (26 in total) we define units of $e$ such that the period of the stability diagram is $2e$.

We collect the results for OS domain size against the binding and charging energy in Figs.~\ref{Fig3}c,d. 
While in QD-SI devices the exchange interaction always stabilizes the screened singlet state, with its size reaching $2e$ for strong binding~\cite{Saldana2022Jan,lee2014spin}, no transition to a singlet state occurs in our SI-QD-SI.
Instead, after an initial decrease of the doublet region,  at large binding the OS state becomes well defined and the domain size becomes approximately constant. This is a sign that the binding energies of the singlet and doublet states vary in the same way. The ground state is always a doublet, only its nature changes as the coupling becomes large.

In QD-SI devices, large $E_\mathrm{c}/\Delta$ stabilizes the screened state by forming a LM in the SI, leading to its size in the GS stability diagram converging to $1e$ independently of the binding strength~\cite{Saldana2022Jan}.
In the SI-QD-SI device, see Fig.~\ref{Fig3}d, the OS domain size grows linearly with $E_c$ at first and then increases sub-linearly until reaching $1e$, at which point one quasiparticle from each SI is bound to the QD spin. Stronger binding is reached by setting the gate voltages more positive, which however prevents large $(E_\mathrm{c}/\Delta)^*$; this results in the clustering of data points into two groups, one at small $(E_\mathrm{c}/\Delta)^*$ for strong binding and one at large $(E_\mathrm{c}/\Delta)^*$ for weak binding, as shown by the color scales in Figs.~\ref{Fig3}c,d.

The size of a domain being the measure of stability of a given ground state, it relates to the excitation energy between the competing ground states. This allows for a qualitative comparison of measured domain sizes to calculated excitation energies shown in Figs.~\ref{Fig3}e,f. The same trends as in Figs.~\ref{Fig3}c,d are indeed observed; a more direct comparison of domain sizes to measured excitation energy on a subset of data shown in Extended Data Fig.~\ref{Ext_Fig4} lends further support. 
The device is found to operate in the regime of comparably large characteristic scales, $U \sim \Delta \sim E_\mathrm{c}$. This complicates the already complex many-body problem as charge fluctuations increase the admixture of configurations with empty or doubly occupied QD. For example, the contribution of an inter-SI singlet (a state with an empty QD and a quasiparticle in each SI forming a singlet, see Extended Data Fig.~\ref{Ext0}) is sizeable in the singlet GS for weak coupling.
Nevertheless, the behaviour in the strong-coupling limit is largely unaffected. 
With increasing $\Gamma$, the LR-singlet is stabilized with respect to the inter-SI singlet because of the different binding mechanisms (exchange vs. superexchange), and the admixture of the inter-SI singlet in the GS is reduced.
By the same token, the doublet GS becomes a pure OS state at large $\Gamma$.
The crossover between the admixture and the pure LR-singlet regime is observed in Fig.~\ref{Fig3}e and f. In the admixture regime, slight variation of parameters that control charge distribution (such as $U$) affects the composition of the eigenstates and has a significant effect on the excitation energy (wider bands in Fig.~\ref{Fig3}e and f for low $\Gamma$). In the regime of ``pure states'' there is no such variability.

LR and OS states of pure character can also be deduced from the linear dependence of the excitation energy on $E_\mathrm{c}/\Delta$ seen in Fig.~\ref{Fig3}f for large $\Gamma/U$. The OS size versus $(E_\mathrm{c}/\Delta)^*$ in Fig.~\ref{Fig3}d displays a linear trend for strong binding, consistent with pure LR and OS states. For small $\Gamma/U$, the larger energy of the admixture singlet produces a steeper dependence of the excitation energy on $E_\mathrm{c}/\Delta$ (Fig.~\ref{Fig3}f).
The admixture explains the sub-linear growth of the OS size ending at $1e$ observed in the experiment in Fig.~\ref{Fig3}d. Model calculations of the OS domain size agree with our interpretations (Extended Data Fig.~\ref{Ext11}). 


Further evidence for the OS state is obtained in the presence of magnetic field that polarizes the LMs. At large field $B$ this leads to qualitative changes in the spectrum as the LR triplet and the quadruplet cross the LR singlet and the OS doublet, respectively, as demonstrated in Fig.~\ref{Fig5}. Spectra for weak and strong binding at two $B$ values in Fig.~\ref{Fig5}a-d illustrate the extraction of the excitation energies that are presented in Fig.~\ref{Fig5}e, while the model calculation results in Fig.~\ref{Fig5}f guide the interpretation. 

In the 1 LM regime (top panels in Fig.~\ref{Fig5}e,f), the GS is a doublet with the LM in the QD. The subgap singlet does not disperse while the doublet splits, increasing the excitation energy proportionally to $g_N/2$, $g_N$ being the g-factor of the QD.
Simultaneously, the triplet states descend in energy with a rate proportional to $g_N/2 + (g_L + g_R)/4$, where $g_L$ and $g_R$ are $g$-factors of the left and right SI, and eventually cross the singlet to become the lowest-energy excitations. Since the triplet decreases in energy at a higher rate than the doublet, the excitation energy starts to decrease at this point. The model predicts another change of slope when the quadruplet becomes the GS, but this regime is not reached in the experiment. 

The situation is different in the 3 LM regime with the overscreened state (bottom panels in Fig.~\ref{Fig5}e,f). At zero field, the OS doublet and LR singlet are separated from the continuum of higher-spin excitations in their corresponding subspaces by the same binding energy $E_B$. Furthermore, in each subspace the excitation energy decreases with increasing field at the same rate relative to the respective GS. This means that the singlet-triplet and doublet-quadruplet crossings occur at roughly the same $B$. The excitation energy thus has a constant linear dependence versus $B$ despite the level crossing. It is proportional to $(g_L + g_R)/4$, owing to an additional polarised quasiparticle spread over both SIs in the states with odd occupancy compared to the ones with an even number of particles. The slight disagreement of the experiment with this simple interpretation is mostly due to unequal $g$ factors. While the model assumes $g_L = g_N = g_R$, we experimentally obtain $g_\mathrm{L}=8.7$, $g_\mathrm{N}=17$, $g_\mathrm{R}=5.9$ for the weak-binding dataset, and $g_\mathrm{L}=8.8$, $g_\mathrm{N}=20$, $g_\mathrm{R}=5.7$ for the strong-binding dataset. These are measured by loading a single LM to the relevant device component by tuning appropriate gate voltages.
In the 1 LM case, stronger binding is expected to renormalize the effective $g$ factor of the doublet state \cite{Delgado_2014}, resulting in a more gradual change of slope of excitation energy. In the 3 LM case stronger binding means purer OS and LR subgap states and with that better agreement to the simple picture of completely polarized states.
Additional data is shown in Extended Data Figs.~\ref{Ext2} and \ref{Ext3}.

In conclusion, we presented strong experimental evidence for the existence of a doublet overscreened subgap state in a SI-QD-SI chain. This state is predicted to exhibit long-range ferromagnetic correlations between the quasiparticles in the SIs.
We find that for typical parameters the spin correlations between Bogoliubov quasiparticles is close to 1/8, half the value of an ideal triplet state, and that in longer SC-QD-... chains this correlation has a slow decay.
These correlations are brought to the micrometer scale in our device by setting $E_\mathrm{c}/\Delta>1$, in contrast to the nanometer scale of YSR chains of magnetic adatoms on $E_\mathrm{c}=0$ superconducting substrates~\cite{Nadj-Perge2014Oct}. Even-length chains of QD-SI singlet dimers should instead lead to long-range antiferromagnetic correlations between the end unpaired elements~\cite{CamposVenuti2006Jun}, not limited by the superconducting coherence length in contrast to Cooper-pair splitters~\cite{Hofstetter2009Oct,Ranni2021Nov}.

Understanding the physics of elementary building blocks is an important first step in the pursuit of large-scale quantum simulators.
For example, a longer odd-length chain can be used to demonstrate the self-similar state of the Wilson-chain vision of the two-channel Kondo model~\cite{Cox1996Nov}, where the central extended doublet is recursively overscreened by the ending SIs (Figs.~\ref{Fig1}a,b). The length dependence of the correlations can be investigated by unloading LMs from the elements of the long chain, effectively shortening its length.

The chain can be mapped to other well-known models by setting its parameters to various special limits. For example, for $\Delta=0$ it maps to the Hubbard chain (triple QD for 3 sites) and, for weak hopping, to the Heisenberg chain. For $E_\mathrm{BL} \neq E_\mathrm{BR}$, it realizes the interacting Su-Schrieffer-Heeger model~\cite{Heeger1988Jul}, and for $E_\mathrm{c}=0$, $\Delta=0$ it simulates the Kondo necklace~\cite{Doniach1977Jul}. Extension to two-dimensional lattices is possible by using nanowire networks~\cite{
OphetVeld2020, VaitiekenasPRL2018, Bottcher2018Nov}, enabling the pursuit of topological spin liquids~\cite{Semeghini2021Dec}.

\section*{Methods}

\textbf{Device fabrication.} A 120-nm wide InAs nanowire with a 7-nm in-situ grown epitaxial Al shell covering three of its facets was deposited with a micromanipulator on a Si/$\mathrm{SiO_2}$ substrate used as a backgate. The device was defined by a series of electron-beam lithography steps. The Al was patterned into two $\approx$300-nm long islands by Transene-D etching. The nanowire was contacted by Ti/Au (5/200 nm) leads following a gentle argon milling to remove the nanowire native oxide. A 5-nm thick layer of $\mathrm{HfO_2}$ was deposited over the device to insulate it from seven Ti/Au top gates deposited thereafter. Gates 1 and 7 were respectively short-circuited to gates $V_\mathrm{L}$ and $V_\mathrm{R}$. 

\textbf{Measurements.} All measurements where performed in an Oxford Triton dilution refrigerator at 30 mK. Higher temperature broadens the $G$ features and erases the even-odd parity alternation of the SIs as expected (see Extended Data Fig.~\ref{Ext9}). $G$ was measured by biasing the source with a lock-in voltage of 5 $\mu$V at a frequency 84.29 Hz on top of $V_\mathrm{sd}$, and recording the lock-in current at the grounded drain. Zero-bias $G$ was measured at -18 $\mu$V to account for an offset in the current amplifier. $B$ was aligned with the nanowire axis to maximize the critical field, $B_\mathrm{c}$. $B_\mathrm{c}$ was estimated at $>1.5$ T. A single QD was achieved by setting $V_\mathrm{3}, V_\mathrm{5}$ to negative values. To achieve left-right symmetry, QD shells with approximately left-right symmetric binding energy were further fine-tuned with $V_\mathrm{L}$ and $V_\mathrm{R}$ until $E_\mathrm{cL}/\Delta_\mathrm{L} \approx E_\mathrm{cR}/\Delta_\mathrm{R}$. To achieve the electron-hole symmetric filling of the QD in Figs.~\ref{Fig3}-\ref{Fig5}, $V_\mathrm{N}$ was fine-tuned until the bottom left and top right parts of the stability diagram were symmetric. Tuning of $E_\mathrm{cL}/\Delta_\mathrm{L}$ and $E_\mathrm{cR}/\Delta_\mathrm{R}$ was achieved by using two auxiliary QDs, one each to the left and right of the left and right SIs. $E_\mathrm{cL,R}$ was reduced when these QDs were put in resonance with the drain and source leads. Though in reality a five element QD-SI-QD-SI-QD chain, the device behaved as a shorter SI-QD-SI chain as intended with the outer QDs set in cotunnelling. We speculate that this was due to low tunnel couplings between the auxiliary QDs and the SIs, and/or due to the auxiliary QDs having even occupation.

\textbf{Asymmetry.} Left-right binding asymmetry precludes the inter-island 
tunnelling of quasiparticles. Experimentally, this is observed as a weaker reduction of $d(\mathrm{OS})$ when one of the binding energies is asymmetrically increased (in comparison to Fig.~\ref{Fig3}c). A left-right asymmetric increase of $E_\mathrm{cL}/\Delta_\mathrm{L}$, $E_\mathrm{cR}/\Delta_\mathrm{R}$ produces a similar stabilization of the doublet domain as for the symmetric $E_\mathrm{c}/\Delta$ case in Fig.~\ref{Fig3}d. However, the stabilization is asymmetric in the $V_\mathrm{L}$, $V_\mathrm{R}$ gate space, with the largest $E_\mathrm{c}/\Delta$ enlarging the doublet domain in the corresponding island gate direction in the stability diagram.

\newcommand{\QD}{\mathrm{QD}}
\newcommand{\SC}{\mathrm{SC}}
\textbf{Model and calculations.}
For calculations in Fig.~\ref{Fig2}c,d and Fig.~\ref{Fig3}e,f, we describe the QD as a single non-degenerate impurity level, as in the single-impurity Anderson model. The SIs are described by the Richardson model, as two sets of equidistant energy levels that represent time-reversal-conjugate pairs in the momentum/orbital space. These are coupled all-to-all by the pairing interaction. This step beyond the BCS mean-field approximation allows for particle number conservation and is required to accurately describe even-odd occupancy effects of the SI with large charging energy $E_\mathrm{c}$. The QD is coupled to all levels of both SIs with the hybridisation terms. The Hamiltonian is
\begin{equation}
    H = H_\QD + \sum_{\beta=L,R} \big( H_\SC^{(\beta)} + H_\mathrm{hyb}^{(\beta)} \big),
\label{eq:H}
\end{equation}
where
\begin{equation*}
\begin{aligned}
    H_\QD &= \varepsilon_\QD \hat{n}_\QD + U \hat{n}_{\QD, \uparrow}  \hat{n}_{\QD, \downarrow} + E_{Z, \QD} \hat{S}_{z,\QD} \\
    &= \frac{U}{2} ( \hat{n}_\QD - \nu )^2 +  E_{Z, \QD} \hat{S}_{z,\QD} + \mathrm{const.}, \\
    H^{(\beta)}_\SC &=  \sum_{i,\sigma}^N \varepsilon_{i} c^\dagger_{i,\sigma,\beta}c_{i,\sigma,\beta} - \alpha_\beta d \sum_{i, j}^N c^\dagger_{i,\uparrow,\beta} c^\dagger_{i,\downarrow,\beta} c_{j,\downarrow,\beta} c_{j,\uparrow,\beta} \\
    & + E_\mathrm{c}^{(\beta)} ( \hat{n}_\SC^{(\beta)} - n_0^{(\beta)} )^2 + E_{Z}^{(\beta)} \hat{S}^{(\beta)}_{z}, \\
    H_\mathrm{hyb}^{(\beta)} &= (v_\beta/\sqrt{N}) \sum_{i,\sigma}^N (c_{i,\sigma,\beta}^\dagger d_\sigma + d_\sigma^\dagger c_{i,\sigma,\beta}) \\
    &+ V_\beta (\hat{n}_\QD - \nu) ( \hat{n}^{(\beta)}_\SC - n_0^{(\beta)}).
\end{aligned} 
\end{equation*}
Here $\varepsilon_\QD$ is the energy level and $U$ the electron-electron repulsion on the QD.
The QD term can be rewritten in terms of $\nu = 1/2 - \varepsilon_\QD/U$, the QD level in units of electron number. 
$d_\sigma$ and $c_{i,\beta,\sigma}$ are the annihilation operators corresponding to the QD and the two SIs labeled by $\beta=L,R$ (left and right). The spin index is $\sigma=\uparrow, \downarrow$. The $N$ SI energy levels $\varepsilon_i$ are spaced by a constant separation $d=2D/N$, where $2D$ is the bandwidth. The levels are coupled all-to-all by a pairing interaction with strength $\alpha$. 
The number operators are $\hat{n}_\QD = \sum_\sigma d_\sigma^\dagger d_\sigma$ for the QD, and $\hat{n}^{(\beta)}_\SC = \sum_{i=1,\sigma}^N c_{i,\sigma,\beta}^\dagger c_{i,\sigma,\beta}$ for each SI; spin operators are $\hat{S}_{z,\QD} = (1/2) (d_\sigma^\uparrow d_\uparrow-d_\downarrow^\dagger d_\downarrow)$, and similarly
for $\hat{S}_z^{(\beta)}$. $E_\mathrm{c}^{(\beta)}$ are the charging energies, with $n_0^{(\beta)}$ the optimal occupation of the SI in units of electron charge.
The SIs are coupled to the QD with the hybridisation strengths $\Gamma_\beta=\pi \rho v_\beta^2$, where $\rho=1/2D$ is the normal-state density of states in each bath. The $V_\beta$ terms describe the capacitive coupling between the QD and SI. We take $D=1$ as the unit of energy.

The results were obtained for $N=100$ levels in each superconductor and we set $\alpha=0.4$, which determines the superconducting gap in the absence of the QD, $\Delta = 0.16$. This value is chosen so that an appropriate number of levels is engaged in the pairing interaction thus minimizing finite-size effects, while also minimizing the finite-bandwidth effect.

The calculations were performed using the density matrix renormalization group method \cite{dmrg} using the iTensor library \cite{itensor}. The conserved quantum numbers are the total number of electrons $n$ and the $z$-component of total spin $S_z$. 
The doublet$\to$singlet excitation energy shown in Fig. \ref{Fig3} is thus given by the energy difference between the ground states of the relevant singlet and doublet sectors $\delta E = E(n=204, S_z=0) - E(n=203, Sz=1/2)$.

\textbf{Single-level approximation.}
The end-to-end spin correlations shown in Fig. \ref{Fig1}b and energies in Fig.~\ref{Fig2}c were obtained using the zero-bandwidth limit of \eqref{eq:H}, i.e. by only considering a single level in each SI. The energies and spin correlations were obtained by direct diagonalization in sectors of fixed $(n, Sz)$.
For the spin correlations, this approach allows us to consider a longer chain of alternating SIs and QDs coupled in series and gives qualitatively correct results.
For the energy spectra, obtaining the complete eigenstate allows for a simple calculation of the space-inversion parity which helps us interpret the nature of the states. Extracting this quantity from the large matrix-product state obtained with the DMRG is much harder. 

The differences in the results obtained with the single-level approximation and the full DMRG model are due to mixing and avoided crossings of the presented states with a continuum of superconducting excitations which are neglected in the single-level approximation.

%

\section*{Acknowledgements}

We thank Felix Von Oppen, Ram\'on Aguado, Martin \v{Z}onda and Andr\'as P\'alyi for useful discussions and Peter Krogstrup for providing nanowire materials. 

\textbf{Funding:} The project received funding from the European Union’s Horizon 2020 research and innovation program under the Marie Sklodowska-Curie grant agreement No.~832645, QuantERA ’SuperTop’ (NN 127900) and FETOpen AndQC (828948). We additionally acknowledge financial support from the SolidQ project by the Novo Nordisk foundation, Carlsberg Foundation, the Independent Research Fund Denmark, the Danish National Research Foundation, Villum Foundation project No.~25310, KU SCIENCE Visiting Scholar program and the Sino-Danish Center. L.~P. and R.~\v{Z}. acknowledge the support from the Slovenian Research Agency (ARRS) under Grant No.~P1-0044, P1-0416, and J1-3008.

\section*{Author contributions statement}

J.C.E.S. conceived the experiments with input from all co-authors. A.V. fabricated the device. J.C.E.S. and A.V. did the experiments. J.C.E.S. performed the data analysis with input from A.V., K.G.R. and J.N.. J.C.E.S., L.P. and R.\v{Z}. interpreted the experimental data. L.P. and R.\v{Z}. did the theoretical analysis. All authors contributed to writing the manuscript.

\section*{Additional information}

Supplementary Information is available for this paper.

The authors declare no competing interests. 

\subsection*{Data availability}

Raw experimental data shown in the paper is available at the data repository ERDA of the University of Copenhagen \href{}{in this link}.

\setcounter{figure}{0}

\renewcommand{\figurename}{Extended Data Figure}


\begin{figure*}[t!]
\centering
\includegraphics[width=0.7\linewidth]{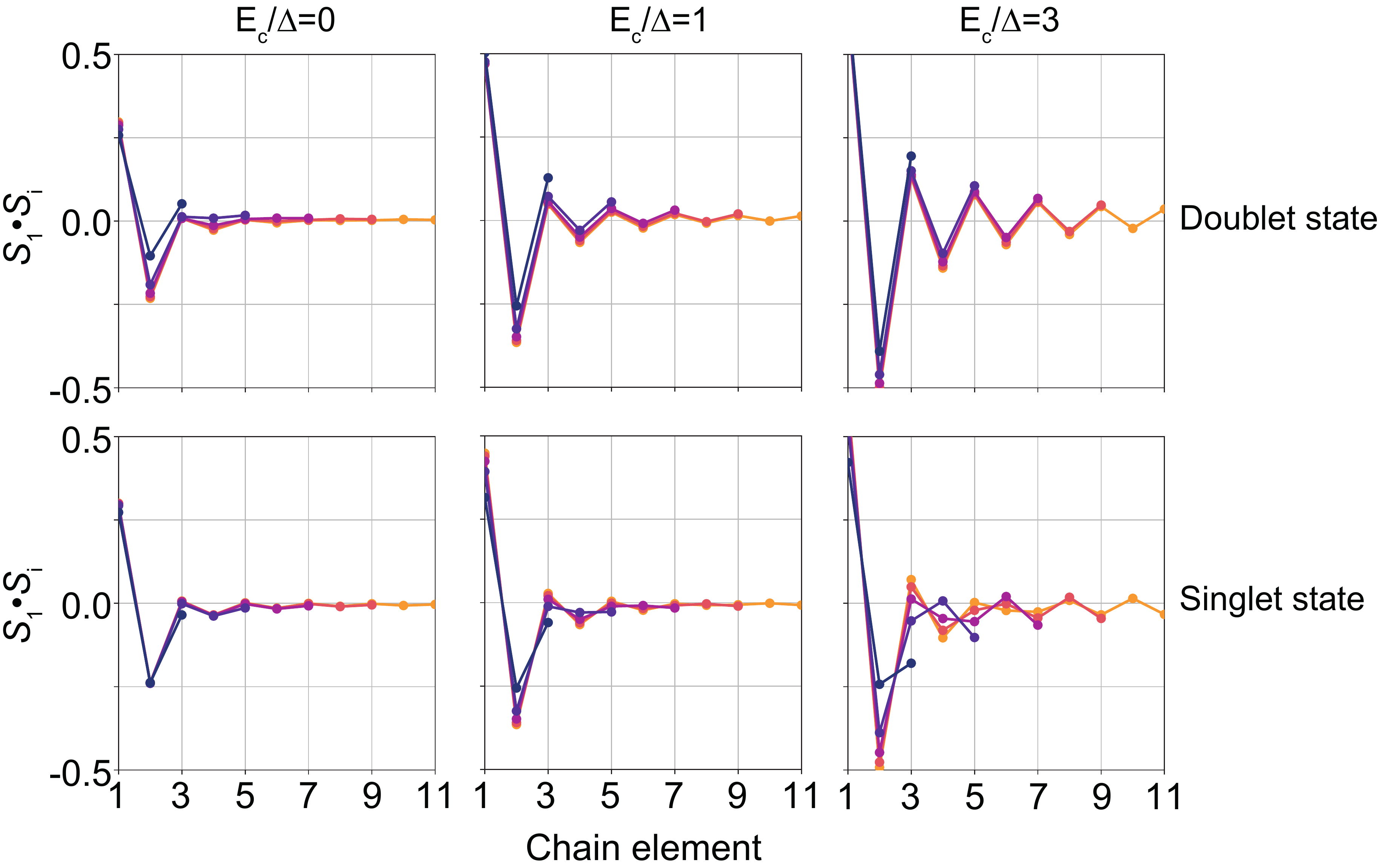}
\caption{\textbf{Spin-spin correlations in alternating chain of superconducting islands and  quantum dots.} Correlations between the first spin, $S_\mathrm{1}$, and the spin in position i, $S_\mathrm{i}$, for $i=1,\ldots,L$, where $L$ is the chain length, calculated with the single-level approximation. Each panel contains the results for chains of different odd lengths (colours), $L=3, 5, 7, 9$ and $11$. Left, central and right panels correspond to $E_\mathrm{c}/\Delta=0$, 1 and 3. Top panels examine the correlations in the doublet state, while bottom ones examine correlations in the singlet state. In the doublet state, the correlations cross zero between adjacent elements for $E_\mathrm{c}/\Delta=1$ and 3, indicating long-range order. The long-range order is lost for $E_\mathrm{c}/\Delta=0$ and for any $E_\mathrm{c}/\Delta$ value in the singlet state.
This figure emphasizes the high degree of self-similarity of spin correlations in the doublet state, demonstrated by the good degree of overlap of the
results for different $L$.
}
\label{Ext9}
\end{figure*}

\begin{figure*}[t!]
\centering
\includegraphics[width=1\linewidth]{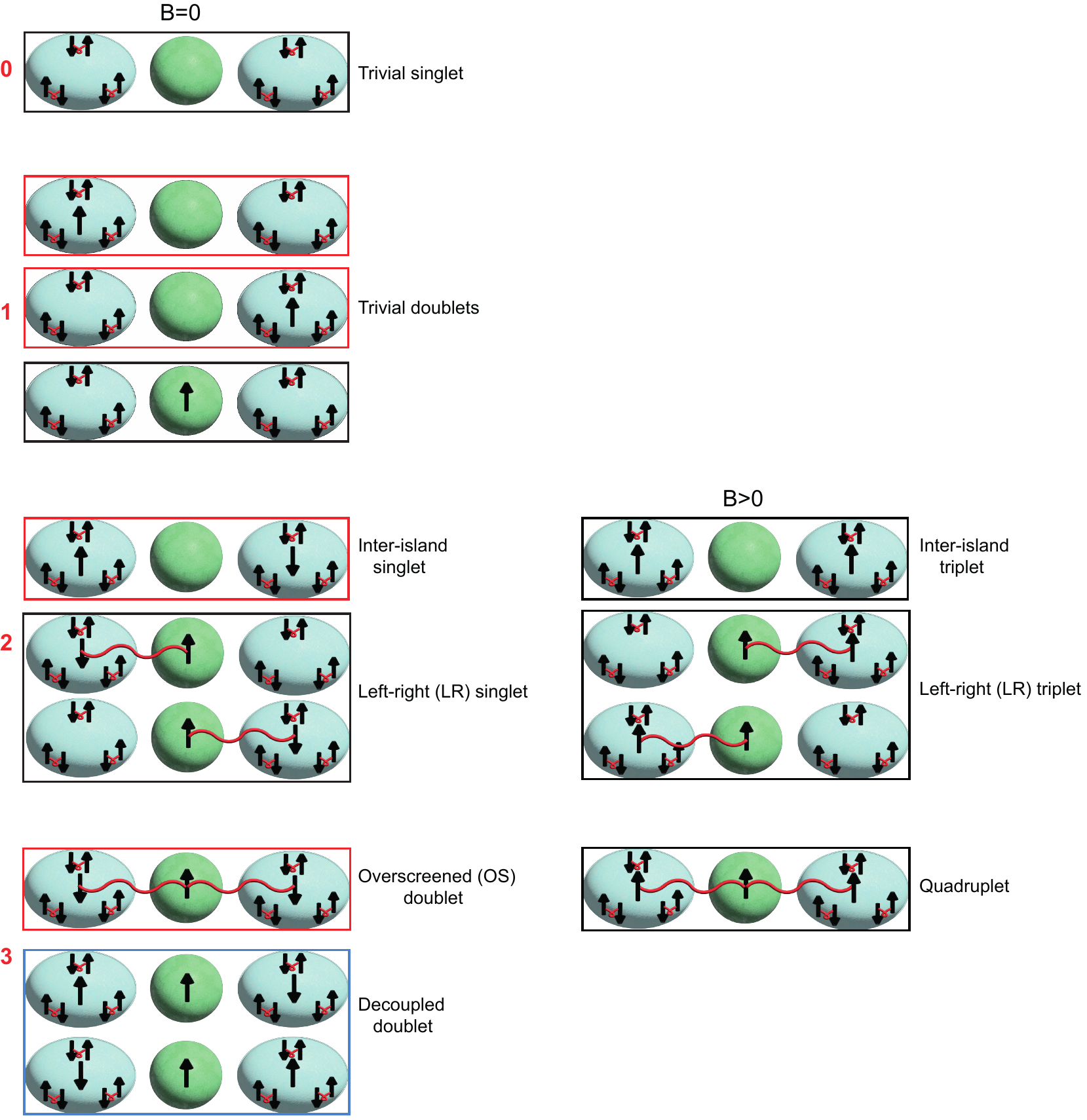}
\caption{\textbf{Ground state for different number of local moments.} Depictions of the wavefunction of the ground state of the SI-QD-SI chain for different total number of LMs. Red frames indicate that the state is the GS only for $E_\mathrm{c}>\Delta$, while blue frames indicate that the state is only the GS for $E_\mathrm{c}<\Delta$. States with black frames can be the GS at any finite $E_\mathrm{c}$. States in the left column can be the GS at $B=0$, and those in the right column may become the GS only at $B>0$. The states in the left column may become the GS at $B>0$ independently of the $E_\mathrm{c}/\Delta$ ratio.}
\label{Ext0}
\end{figure*}

\begin{figure*}[t!]
\centering
\includegraphics[width=0.75\linewidth]{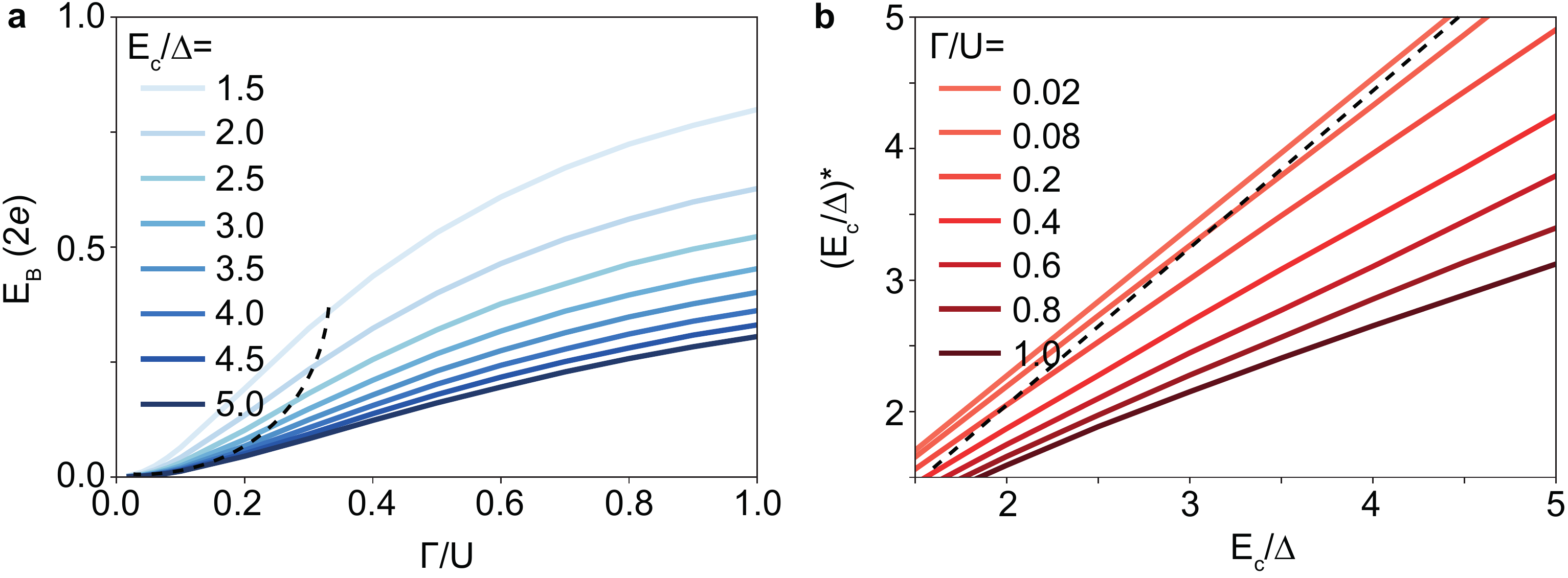}
\caption{\textbf{Relation between parameters extracted from stability diagrams and model parameters.} \textbf{a},\textbf{b} Calculations using the DMRG of the relations between (\textbf{a}) the binding energy $E_\mathrm{B}$ and the hybridisation strength $\Gamma$ and (\textbf{b}) the effective ratio $(E_\mathrm{c}/\Delta)^*$ (extracted from the stability diagram, as explained in the caption to Fig.~\ref{Fig3}) and the ratio $E_\mathrm{c}/\Delta$ (raw model parameters). These relations are linear to a good approximation in the parameter regime of the experiment (corresponding approximately to the left of the dashed lines), supporting the comparison of Figs.~\ref{Fig3}c,d with Figs.~\ref{Fig3}e,f.
}
\label{Ext10}
\end{figure*}

\begin{figure*}[t!]
\centering
\includegraphics[width=1\linewidth]{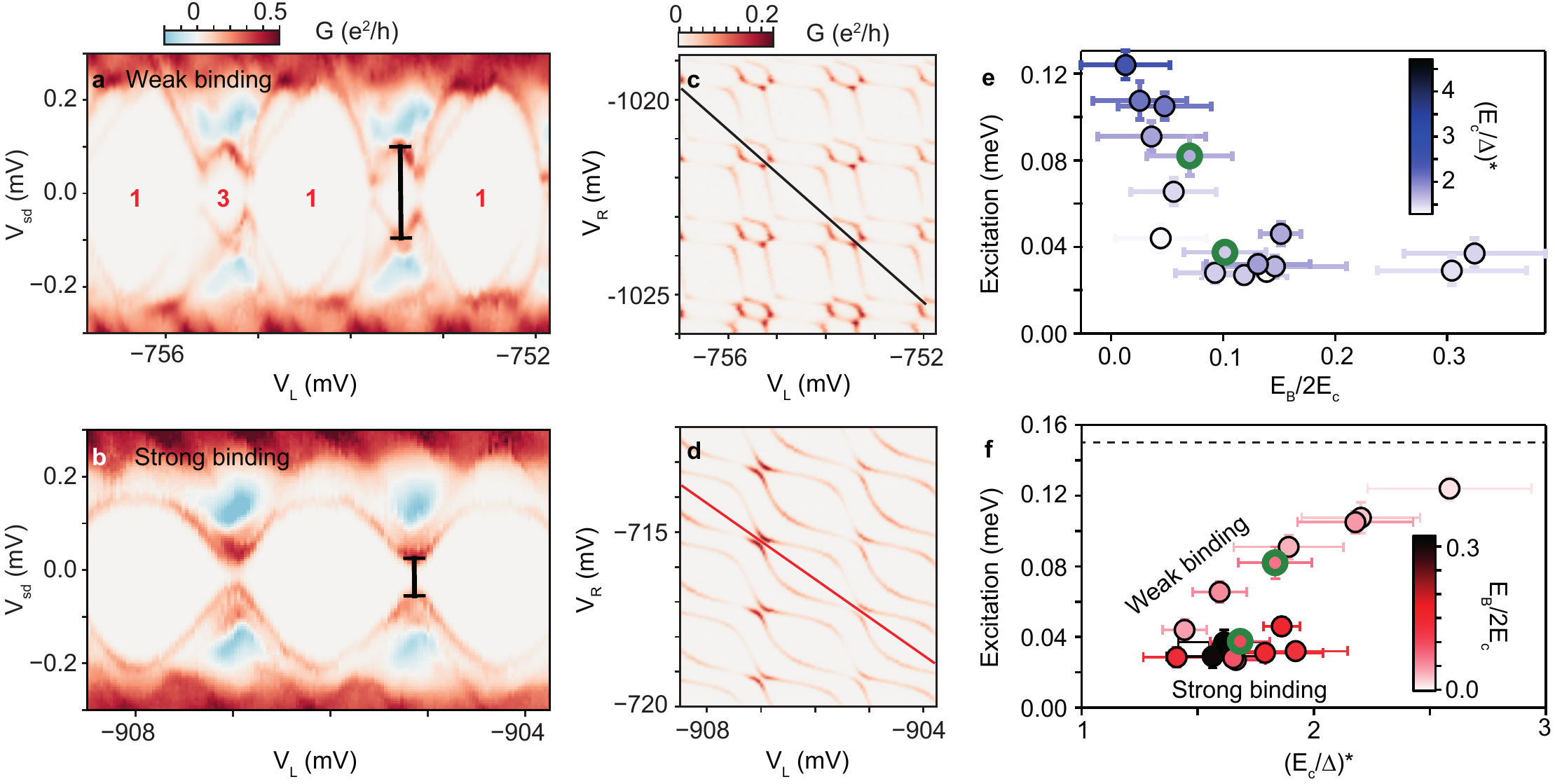}
\caption{\textbf{Overscreened doublet to left-right singlet excitations.} \textbf{a,b} Examples of bias spectra for \textbf{(a)} weak and \textbf{(b)} strong $E_\mathrm{B}$. \textbf{c,d} Zero-bias $G$ stability diagrams with 1 LM fixed in the QD. Solid lines indicate the $V_\mathrm{L}$, $V_\mathrm{R}$ trajectories in \textbf{(a)},\textbf{(b)}. In these trajectories, the number of LMs in the SIs is varied between 2 and 0 in alternation. Red numbers \textbf{(a)},\textbf{(b)} indicate the total number of LMs in the GS. \textbf{e}, \textbf{f} Compilation of doublet$\to$singlet excitation energies in the middle of the OS GS sector for 16 QD shells versus (\textbf{e}) $E_\mathrm{B}/2E_\mathrm{c}$ and (\textbf{f}) $(E_\mathrm{c}/\Delta)^*$. Left-right differences in $E_\mathrm{B}/2E_\mathrm{c}$ and $(E_\mathrm{c}/\Delta)^*$ are less than 30\% and 25\%, respectively. To avoid zero-bias offset issues, the excitation energies are measured as the average of the sum of the addition and removal energies. The sum is given by vertical bars in the examples in \textbf{(a)}, \textbf{(b)}. Green circles are the excitation energy values extracted from these examples. Error bars are based on the full width at half-maximum of the conductance lines used in the extraction of the data. In (\textbf{f}), the weak-binding data is expected to fully saturate at an excitation energy equal to $U/2=0.15$ meV (indicated by a dashed line), where $U$ is measured from Coulomb-diamonds spectroscopy.
}
\label{Ext_Fig4}
\end{figure*}

\begin{figure*}[t!]
\centering
\includegraphics[width=0.9\linewidth]{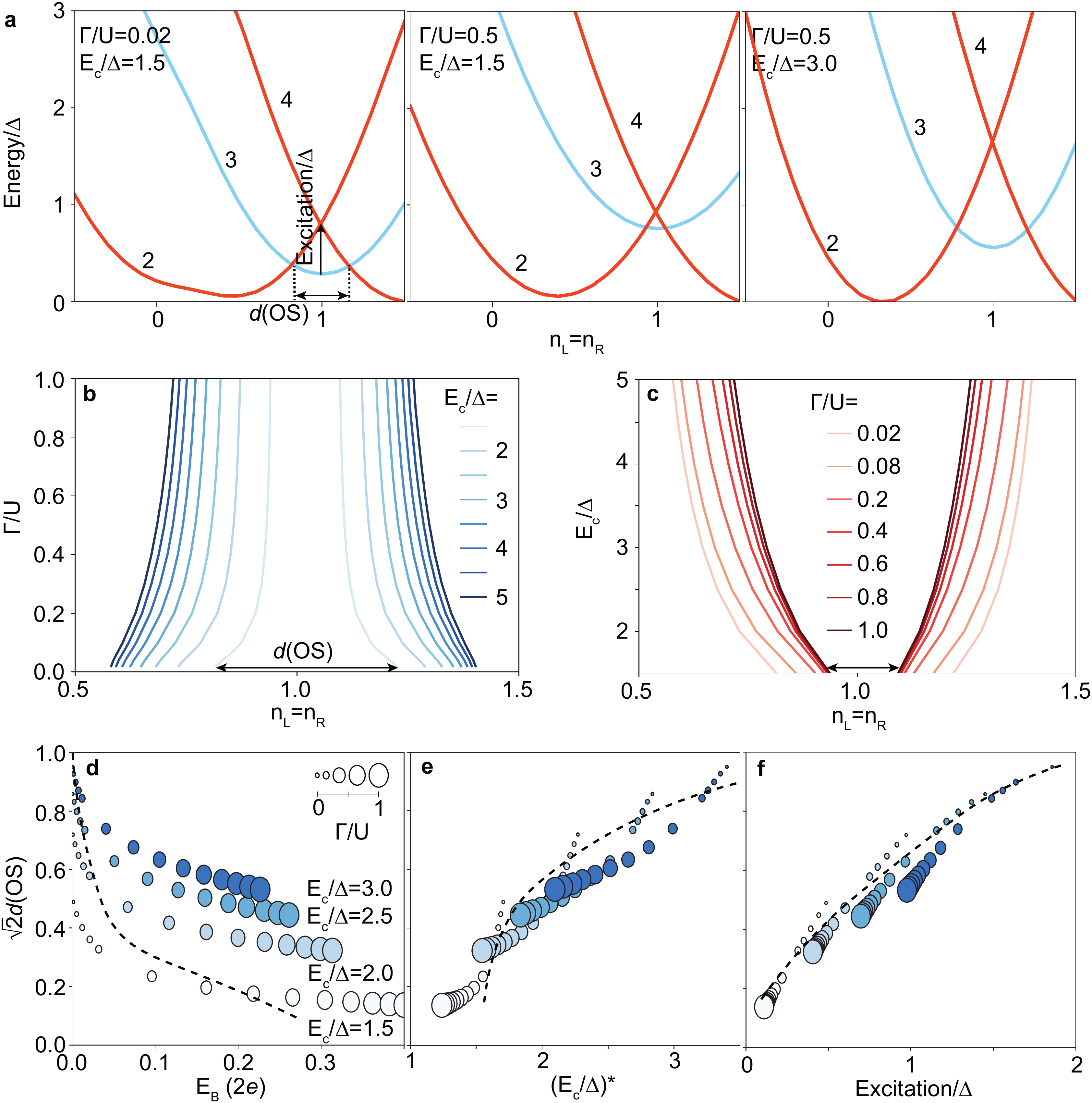}
\caption{\textbf{Calculations of the overscreened doublet size.} \textbf{a} Calculated charge parabolas versus rigid shift of the gate-induced charge in the two SIs $n_\mathrm{L}=n_\mathrm{R}$, for fixed gate-induced charge in the QD $\nu=1$, corresponding to the direction of the diagonal dashed line in Fig.~\ref{Fig3}b. The parabolas are tagged by the total charge in the system referenced to an even integer number of electrons in each of the SIs. The size of the overscreened doublet, $d(\mathrm{OS})$, and the doublet to singlet excitation energy are indicated by horizontal and vertical arrows. An increase in $\Gamma/U$ or a reduction in $E_\mathrm{c}/\Delta$ reduce the excitation energy and $d(\mathrm{OS})$. \textbf{b}, \textbf{c} Parabola crossings versus (\textbf{b}) $\Gamma/U$ for different $E_\mathrm{c}/\Delta$ and versus (\textbf{c}) $E_\mathrm{c}/\Delta$ for different $\Gamma/U$. $d(\mathrm{OS})$ corresponds to the distance between the curves, indicated by horizontal arrows in the innermost pair of curves. \textbf{d}, \textbf{e}, \textbf{f} Calculated $\sqrt{2}d(\mathrm{OS})$ versus (\textbf{d}) $E_\mathrm{B}$, (\textbf{e}) $(E_\mathrm{c}/\Delta)^*$, and (\textbf{f}) excitation energy normalized by $\Delta$. The parameter region encompassed by the experiment corresponds approximately to the left of the dashed lines. $\sqrt{2}d(\mathrm{OS})$ grows approximately linearly with the excitation energy, showing that the former is an appropriate measure of the latter and enabling the comparisons made in Fig.~\ref{Fig3} between these two quantities.  
}
\label{Ext11}
\end{figure*}

\begin{figure*}[t!]
\centering
\includegraphics[width=0.9\linewidth]{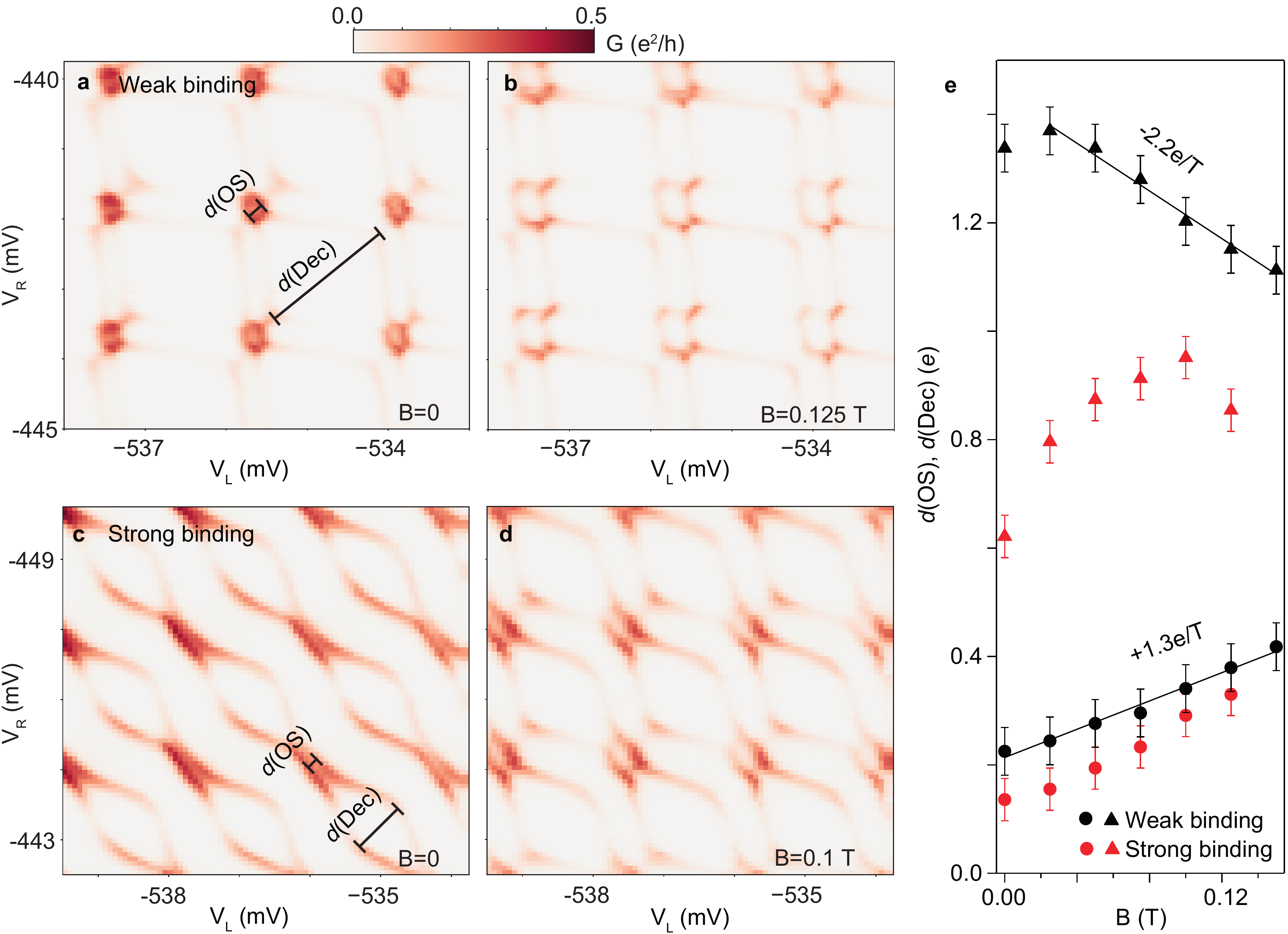}
\caption{\textbf{Evolution of the stability diagram in a magnetic field.} \textbf{a-d} Zero-bias conductance stability diagrams recorded at different $B$ indicated on each plot. The device parameters are identical to those of Fig.~\ref{Fig5}, i.e. (\textbf{a,b}) $E_\mathrm{B}/2E_\mathrm{c}=0.04$ (weak binding), $E_\mathrm{c}/\Delta=1.45$ and (\textbf{c,d}) $E_\mathrm{B}/2E_\mathrm{c}=0.32$ (strong binding), $E_\mathrm{c}/\Delta=1.65$. \textbf{e} $B$ dependence of the diagonal sizes in the 3 LM (circles) and 1 LM (triangles) sectors extracted from stability diagrams for weak (black symbols) and strong binding (red symbols). The sizes at $B=0$ are given by the bars in (\textbf{a}), (\textbf{b}). Lines are fits to the weak-binding data, with slopes indicated above each line. The data is qualitatively consistent with the Zeeman shifts of the excitation energies shown in Fig.~\ref{Fig5}e.}
\label{Ext2}
\end{figure*}

\begin{figure*}[t!]
\centering
\includegraphics[width=0.6\linewidth]{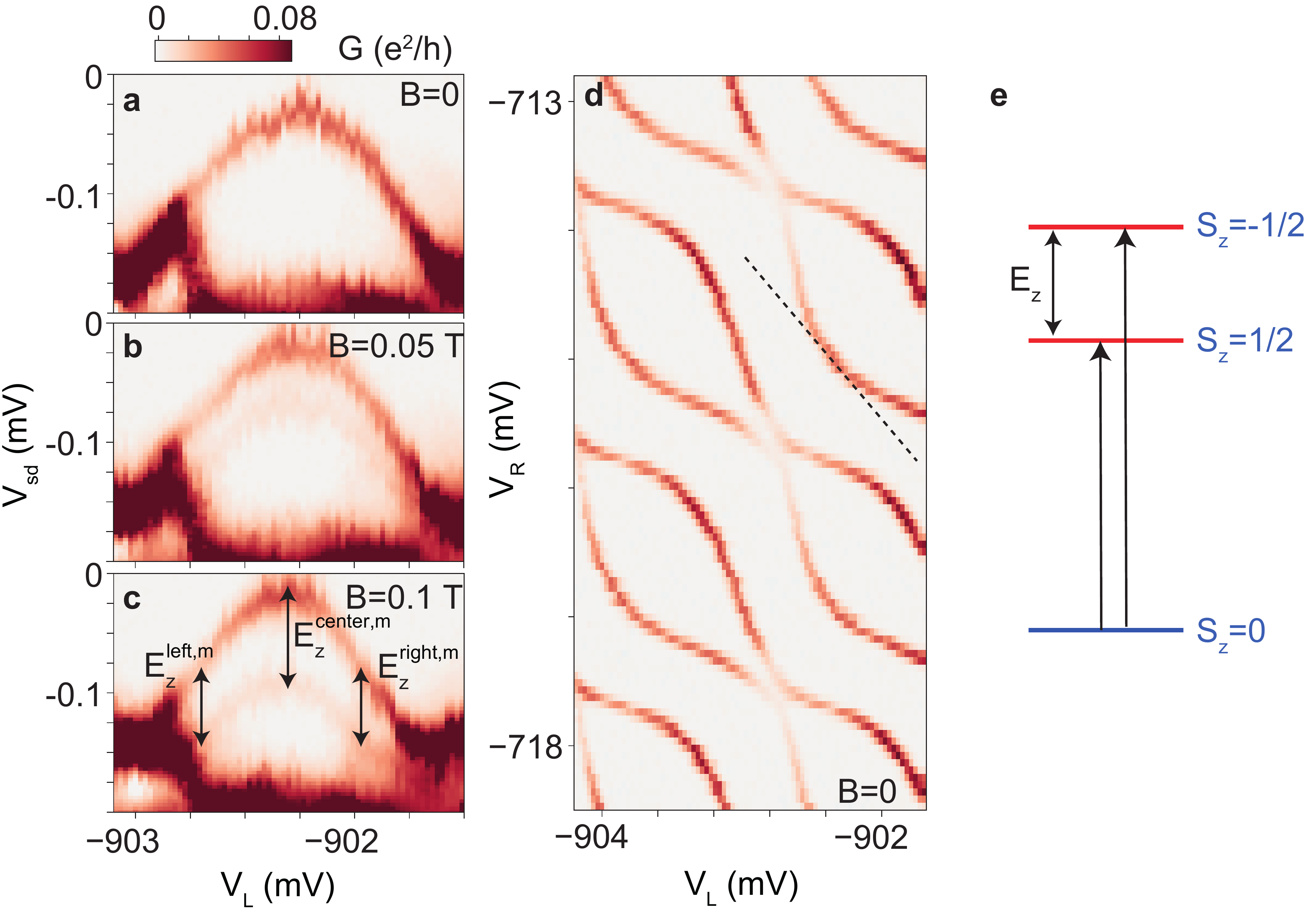}
\caption{\textbf{Zeeman splitting of left-right singlet to doublet excitations.} \textbf{a-c} Magnetic field, $B$, dependence of the $G$ versus source-drain bias, $V_\mathrm{sd}$, with the gates $V_\mathrm{L}$ and $V_\mathrm{R}$ swept along the dashed line in the zero-bias $G$ stability diagram in (\textbf{d}). For simplicity, only $V_\mathrm{L}$ is indicated. The QD is half-filled, left-right binding energies are approximately symmetric and strong ($E_\mathrm{B}/2E_\mathrm{c}=0.15>\mathcal{E}$), and $E_\mathrm{cL}/\Delta_\mathrm{L} \approx E_\mathrm{cR}/\Delta_\mathrm{R} = 1.9$. The curved line in the measurement represents singlet$\to$doublet excitations with gate-dependent state mixing, with the doublet state splitting into spin-up and spin-down components as $B$ is increased. In (\textbf{c}), Zeeman energies at three spots in the split curve are indicated by double-headed arrows; they highlight that the splitting is larger at the center of the curve. The color scale is saturated to highlight the faint excitations. The $g$ factors of the three device components are fairly symmetric: $g_\mathrm{L}=12.6$, $g_\mathrm{N}=15$ and $g_\mathrm{R}=11$. The Zeeman splitting at the center is $E_\mathrm{z}^\mathrm{center,m}=85 \pm 18$ $\mu$eV, which matches well the Zeeman splitting of the trivial doublet excitation $E_\mathrm{z}^\mathrm{center}=g_\mathrm{N}\mu_\mathrm{B} B=87$ $\mu$eV as expected. At the left and right sides, the Zeeman splitting is $E_\mathrm{z}^\mathrm{left,m}=E_\mathrm{z}^\mathrm{right,m}=67 \pm 16$ $\mu$eV. The expected Zeeman splittings three-quarters of the way (in gate voltage) to the OS doublet are $E_\mathrm{z}^\mathrm{left}=E_\mathrm{z}^\mathrm{right}=0.5g_\mathrm{N}\mu_\mathrm{B}B+0.5(n_\mathrm{L}g_\mathrm{L}+n_\mathrm{R}g_\mathrm{R})\mu_\mathrm{B} B$, where $n_\mathrm{L}$ and $n_\mathrm{R}$ are the gate-induced charges in the left and right SIs. These evaluate as $E_\mathrm{z}^\mathrm{left}=54$ $\mu$eV for $n_\mathrm{L}=0.75$, $n_\mathrm{R}=0.25$, and $E_\mathrm{z}^\mathrm{right}=68$ $\mu$eV for $n_\mathrm{L}=0.25$, $n_\mathrm{R}=0.75$, matching within error bars the measurements. \textbf{e} Schematics of singlet$\to$doublet excitations (arrows). The energy difference between the excitations provides a measure of the Zeeman energy.}
\label{Ext3}
\end{figure*}

\end{document}